# Change of Cr atoms distribution in Fe$_{85}$Cr$_{15}$ alloy caused by 250 keV He$^+$ ion irradiation to different doses

S. M. Dubiel[1*] and J. Żukrowski[2]

[1]AGH University of Science and Technology, Faculty of Physics and Applied Computer Science, al. Adama Mickiewicza 30, 30-059 Kraków, Poland

[2]AGH University of Science and Technology, Academic Centre for Materials and Nanotechnology and Faculty of Physics and Applied Computer Science, al. A. Mickiewicza 30, 30-059 Kraków, Poland

Redistribution of Cr atoms in a Fe$_{85}$Cr$_{15}$ alloy caused by its irradiation with 250 keV He$^+$ ions to different doses, $D$=8·10$^{16}$, 16·10$^{16}$ and 48·10$^{16}$ ions/cm$^2$ was investigated by means of conversion electrons Mössbauer spectroscopy. The redistribution was expressed in terms of the Warren-Cowley short-range order parameters $\alpha_1$, $\alpha_2$ and $\alpha_{12}$ pertaining to the first (1NN), second (2NN) and both i.e. 1NN+2NN shells, respectively. Clear evidence was found, both for non-irradiated and irradiated samples that the actual distribution of Cr atoms is characteristic of the shell, and for a given shell it depends on the irradiation dose. In particular, $\alpha_1$ is positive, hence indicates an under population of Cr atoms in 1NN with respect to the random case, $\alpha_2$ is negative, giving evidence thereby that 2NN is overpopulated by Cr atoms, and $\alpha_{12}$ is weakly positive. Under the applied irradiation the number of Cr atoms in both neighbor shells decreased signifying thereby a clustering of Cr atoms. The underlying decrease of Cr concentration within the 1NN-2NN volume around the probe Fe atoms was estimated at 1.5 at% ranging between 2.1 for the lowest and 0.8 at% for the highest dose.

[*]Corresponding author: Stanislaw.Dubiel@fis.agh.edu.pl

# 1. Introduction

High chromium ferritic/martensitic (F/M) steels are manufactured based on Fe-Cr alloys. Their industrial and technological importance follows from their very good useful properties like swelling, high temperature corrosion, creep resistance and other [1,2]. Consequently, they have been regarded as appropriate construction materials to be applied for the new generation of nuclear power facilities such as generation IV fission reactors and fusion reactors as well as for other technologically important plants like high power spallation targets [3–5]. In particular, they are used for a construction of such devices as fuel cladding, container of the spallation target or primary vessel. These devices work at service not only at elevated temperatures but also under irradiation conditions. In these circumstances, the materials undergo irradiation damage that can seriously degrade their properties limiting thereby the lifetime of nuclear reactors and other devices manufactured therefrom. On the lattice scale, the radiation produces lattice defects and a redistribution of Fe/Cr atoms. The latter leads to a short-range ordering (SRO) or causes a phase decomposition into Fe-rich ($\alpha$) and Cr-rich ($\alpha'$) phases. Both these effects result, among other, in an enhancement of embrittlement. A better knowledge of the effect of irradiation on the useful properties of F/M steels and underlying mechanisms is an important issue as it may help to significantly improve nuclear materials properties, hence to extend the durability of devices constructed therefrom. Fe–Cr alloys, the major component of F/M steels, have been regarded as model alloys for investigations of both physical and technological properties ([6] and references therein). In this paper the effect of $^4He^+$ irradiation to different doses with ions of 250 keV on a model (EFDA/EURATOM) $Fe_{85}Cr_{15}$ alloy was studied with the conversion electrons Mössbauer spectroscopy (CEMS). The issue is of a practical importance, as a production of helium occurs during exposure of the devices to proton and/or neutron irradiation [1]. Its presence is known to have a deteriorating effect on mechanical properties of materials. In particular, it lowers the critical stress for inter granular structure and also it may induce a severe decrease of the fracture toughness [7]. Therefore, the understanding not only of the radiation-induced damage but also of the effect of helium on mechanical properties of F/M steels is one of the important topics to be further studied. Numerous studies devoted to the issue have been already done both on steels as well as on the model Fe-Cr alloys. Concerning the latter irradiations with different particles/ions ($Fe^+$, $He^+$, neutrons) e. g. [8-11] were performed and irradiated samples were investigated with different techniques (SANS, TEM, ATP, nanoindentation e. g. [8-13] to reveal irradiation-caused changes in the microstructure e. g. [9,10], in the distribution of Cr atoms e. g. [8, 10,11] or in hardening e. g. [13]. The Mössbauer spectroscopy (MS), although relevant to investigate the irradiation-induced effects in Fe-Cr alloys, was applied very seldom. Its relevance to study the issue follows from a high sensitivity of hyperfine parameters, and, first of all that of the hyperfine magnetic field, $B$, to a presence of foreign atoms in the vicinity of the probe $^{57}$Fe atoms. In general, $B$ can "see" the presence of the foreign atoms situated in the first (1NN) and in the second (2NN) neighbor-shells. For example, one Cr atom present in 1NN changes $B$ by ~3.1T (~9%), and by ~2.1T (~6%) if situated in 2NN. The effect of Cr atom located in 3NN is ~0.3T. Consequently, the latter is usually neglected in the analysis of the Mössbauer spectra. As the influence of Cr atoms on $B$ is additive, measurements of $B$ permit counting Cr atoms within the 1NN-2NN volume around the probe Fe atoms, hence to trace changes in their distribution caused by various external factors like heat treatment, irradiation, etc. This kind of data is of a great importance as it can be used for verification of various theoretical models/approaches/simulations used in order to get a better understanding of the physics underlying such processes like phase-decomposition, irradiation-induced clustering, which are responsible for deterioration of useful properties of materials manufactured based on Fe-Cr alloys. In these calculations interactions between atoms being the first and the second neighbors are usually taken into account. It is worth noting that contrary to other microscopic methods like ATP and TEM, MS gives information on the whole sample with a resolution of a unit cell. More details can be found elsewhere [14-19].

**2. Samples, spectra measurement and analysis**

Samples investigated in this study were prepared from a model EFDA/EURATOM master $Fe_{85}Cr_{15}$ alloy fabricated in 2007. It was delivered in the form of bar 10.9 mm in diameter, in a re-crystallized state after cold reduction of 70% and then heat-treated for 1h under pure argon flow at 850°C followed by air cooling. For the present study, a slice ~1 mm thick was cut off from the bar using a diamond saw, and it was subsequently cold-rolled (CR) down to a final thickness of ~25-30 µm. For the irradiation with 250 keV $^4He^+$ ions at the JANNUS multi-ion beam irradiation platform at Centre National de la Recherche Scientifique, Saclay, France, samples in form of ~25 mm rectangles were used. They were irradiated with a flux of $1.0(1) \cdot 10^{13} He^+ cm^{-2} s^{-1}$ to the dose of $8 \cdot 10^{16}$, $16 \cdot 10^{16}$ and $48 \cdot 10^{16}$ $^4He^+ \cdot cm^{-2}$ or 5, 10 and 30 dpa, respectively, The irradiation area had a diameter of 20 mm, and the temperature was stabilized at 20±2°C (however, strong temperature gradient of 30°C between samples and thermocouple occurred). An exemplary ion range and concentration profile is shown in Fig. 1. It clearly demonstrates that the zone in which the implanted $He^+$ had accumulated was not accessible to the CEMS measurements described in this paper - see below. In other words, the measured spectra contain information on the ballistic effect of He ions passing through the pre layer of their irradiated sides whereas they do not give information on the influence of the presence of these ions.

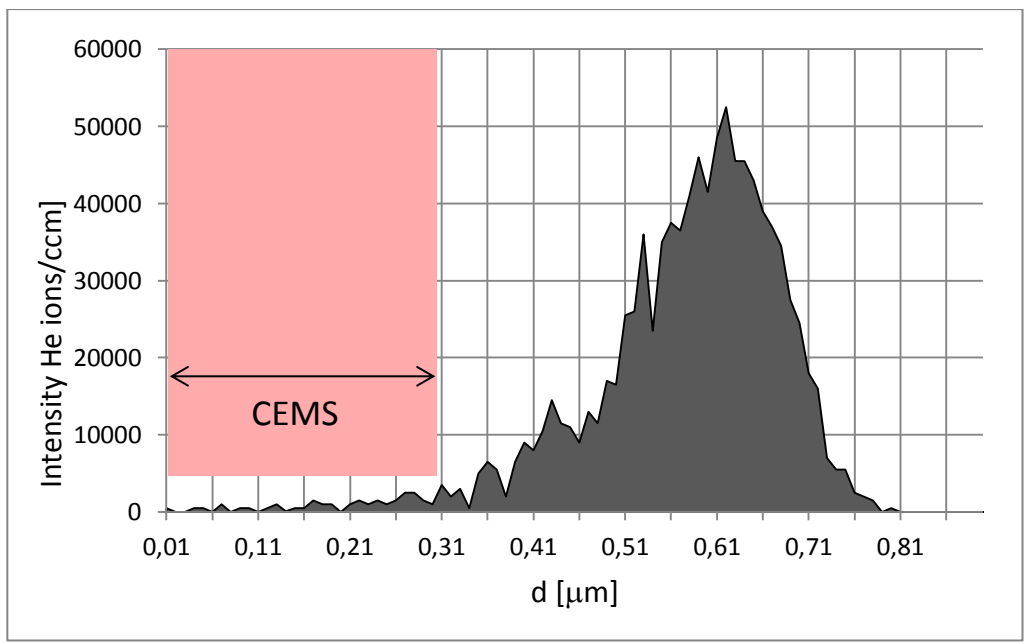

Fig. 1. He-concentration profile as measured for the $Fe_{85}Cr_{15}$ sample irradiated to the dose of 5 dpa with 250 keV $He^+$. The pre surface layer measured with CEMS is marked by the vertical stripe. The average range of the implanted ions is equal to 0.58 µm and the peak concentration ~5 at%He.

$^{57}$Fe-site Mössbauer spectra were measured at room temperature (RT) recording conversion electrons (CEMS mode) in a backscattering geometry using a conventional constant acceleration spectrometer and a $^{57}$Co(Rh) source of 14.4 keV gamma-rays with a nominal activity of 3.7 GBq. This means that they contain information on a pre surface layer whose thickness is less than ~0.3 μm. Samples were placed in a proportional gas flow counter with a He/methane mixture as counting gas. The spectra recorded at room temperature both on the irradiated (IR) as well as on the non-irradiated (NIR) sides are displayed in Fig. 2. They were analyzed applying a two-shell model i.e. assuming spectral parameters viz. the hyperfine field, *B*, and the isomer shift, *IS*, are sensitive to Cr atoms present within the first (1NN) and the second (2NN) neighbor shells. Furthermore, based on results reported elsewhere [8,9], the effect of the presence of Cr atoms in the 1NN-2NN volume of the $^{57}$Fe probe nuclei on *B* and *IS* was assumed to be additive. In other words the relationship $X(n_1,n_2) = X(0,0) + n_1 \cdot \Delta X_1 + n_2 \cdot \Delta X_2$, where $X = B$ or $IS$, $\Delta X_i$ is a change of *B* or *IS* due to one Cr atom situated in 1NN (*i=1*) or in 2NN (*i=2*), was assumed to hold. The number of Cr atoms in 1NN is indicated by $n_1$, and that in 2NN by $n_2$. Based on the binomial distribution, seventeen most probable atomic configurations, $(n_1,n_2)$, were chosen to fulfill the condition $\sum_{n_1,n_2} P(n_1,n_2) \geq 0.99$ i.e. the spectra were analyzed in terms of seventeen sextets. However, in the fitting procedure their probabilities, $P(n_1,n_2)$, were treated as free parameters (their starting values were those calculated from the binomial distribution). All spectral parameters involved in the fitting procedure i.e. $X(0,0)$, $\Delta X_i$, line widths of individual sextets *G1*, *G2* and *G3* and their relative intensities (Clebsch-Gordan coefficients) *C2* and *C3* were treated as free (*C1*=1). Very good fits (in terms of $\chi^2$) were obtained and the best-fit values of the spectral parameters are presented in Table 1. Noteworthy, these values are in a very good agreement with the corresponding ones obtained previously for Fe-Cr alloys [14-19].

The knowledge of the atomic configurations, $(n_1,n_2)$, and their probabilities, $P(n_1,n_2)$, permitted next to determine the average number of Cr atoms in 1NN, $<n_1> = \sum_{n_1,n_2} n_1 P(n_1,n_2)$, in the second, $<n_2> = \sum_{n_1,n_2} n_2 P(n_1,n_2)$, and in both shells, $<n_{12}> = \sum_{n_1,n_2} (n_1+n_2) P(n_1,n_2)$.

Knowing the values of $<n_1>$, $<n_2>$, and $<n_{12}>$ corresponding SRO parameters, $\alpha_1$, $\alpha_2$, and $\alpha_{12}$ could have been calculated as outlined in Section 3.

Table 1

Best-fit values of spectral parameters obtained from the spectra recorded on the non-irradiated (NIR) and irradiated (IR) sides. Meaning of the parameters is given in the text. The values of *IS(0,0)* are relative to the $^{57}$Co(Rh) source.

| Sample | Dose ($10^{16}$/cm$^2$) | B(0,0) [T] | $\Delta B_1$ [T] | $\Delta B_2$ [T] | IS(0,0) [mm/s] | $\Delta IS_1$ [mm/s] | $\Delta IS_2$ [mm/s] | G1/2 [mm/s] | G2/2 [mm/s] | G3/2 [mm/s] | C2/C3 |
|---|---|---|---|---|---|---|---|---|---|---|---|
| Fe$_{85}$Cr$_{15}$ | 8/NIR | 34.04(3) | -3.08(6) | -1.98(3) | -.093(1) | -.018(1) | -.010(1) | .180(5) | .159(4) | .145(2) | 3.3(1) |
|  | 16/NIR | 33.79(3) | -3.17(6) | -2.02(6) | -.099(2) | -.018(2) | -.014(3) | .172(5) | .153(5) | .123(2) | 2.9(1) |
|  | 48/NIR | 33.79(3) | -3.09(5) | -2.07(4) | -.098(2) | -.024(1) | -.011(2) | .172(6) | .143(4) | .127(3) | 3.3(1) |
| Fe$_{85}$Cr$_{15}$ | 8/IR | 33.99(3) | -3.04(4) | -1.98(3) | -.092(2) | -.022(1) | -.015(1) | .161(5) | .137(4) | .128(2) | 3.1(1) |
|  | 16/IR | 33.88(3) | -3.04(5) | -1.91(4) | -.097(2) | -.020(2) | -.014(2) | .161(5) | .126(4) | .120(2) | 3.0(1) |
|  | 48/IR | 33.76(3) | -3.06(4) | -1.98(3) | -.098(2) | -.020(1) | -.012(2) | .165(5) | .139(4) | .120(2) | 3.5(1) |

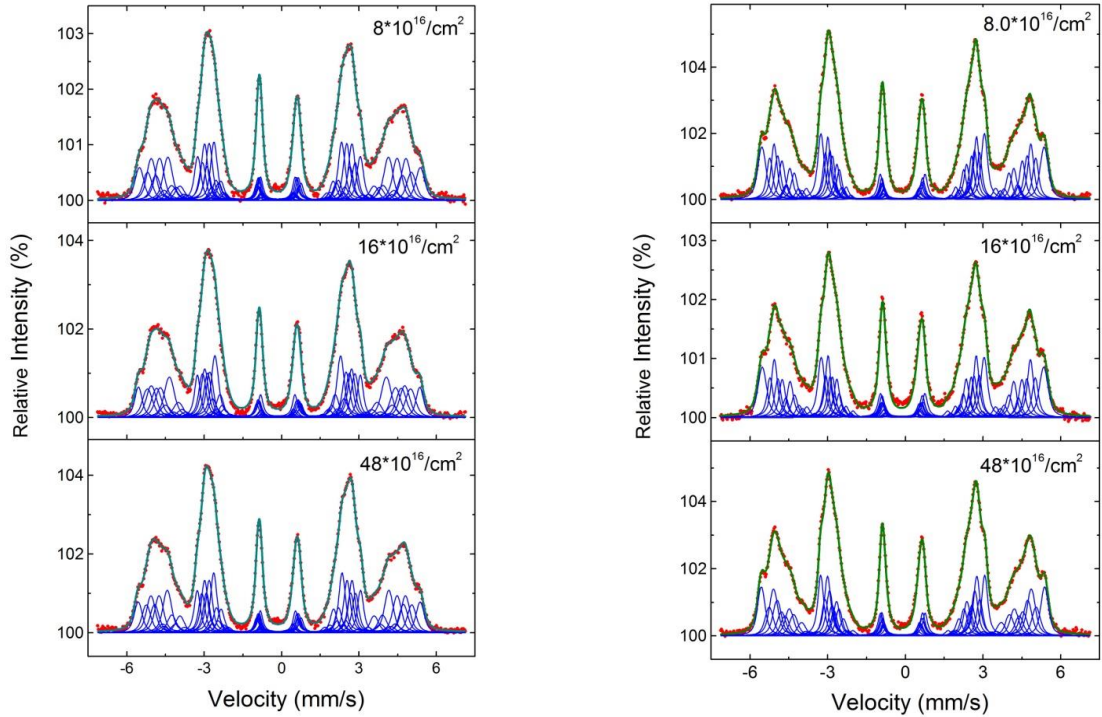

Fig. 2. $^{57}$Fe Mössbauer spectra recorded at RT on the non-irradiated sides of the samples (left panel) and those recorded on the irradiated sides of them (right panel). Subspectra corresponding to particular atomic configurations taken into account in the fitting procedure are indicated.

## 3. Results and discussion
### 3.1. Short-range order (SRO) parameters

A distribution of Cr atoms in the iron matrix can be quantitatively described in terms of Warren-Cowley short-range order (SRO) parameters, $\alpha_i$. The method applied in the present study i.e. the Mössbauer spectroscopy enables determination of the SRO-parameters for the first, $\alpha_1$, and for the second, $\alpha_2$, nearest-neighbor shells, separately. Knowing both of them, one can also calculate the SRO-parameter for both shells, $\alpha_{12}$, as weighted average. The knowledge of the SRO-parameters makes it possible to discuss the distribution of atoms in terms of their ordering or clustering (anti clustering). The values of $\alpha_k$ ($k$=1,2,12) were calculated based on the following equation:

$$\alpha_k = 1 - \frac{<n_k>}{<n_{ok}>} \qquad (1)$$

Where $<n_k>$ is the actual number of Cr atoms in the $k$-th near-neighbor shell, while $<n_{ok}>$ is the number of Cr atoms in the $k$-th near-neighbor shell calculated assuming their distribution is random i.e. $<n_{01}>$=0.08$x$, $<n_{02}>$=0.06$x$, and $<n_{01}+n_{02}>$=0.14$x$.

### 3.1.1. Non-irradiated (NIR) side

Based on the analysis of the spectra recorded on the NIR-sides of the IR-samples the SRO-parameters $\alpha_1$, $\alpha_2$ and $\alpha_{12}$ were determined and they are displayed in Fig. 3.

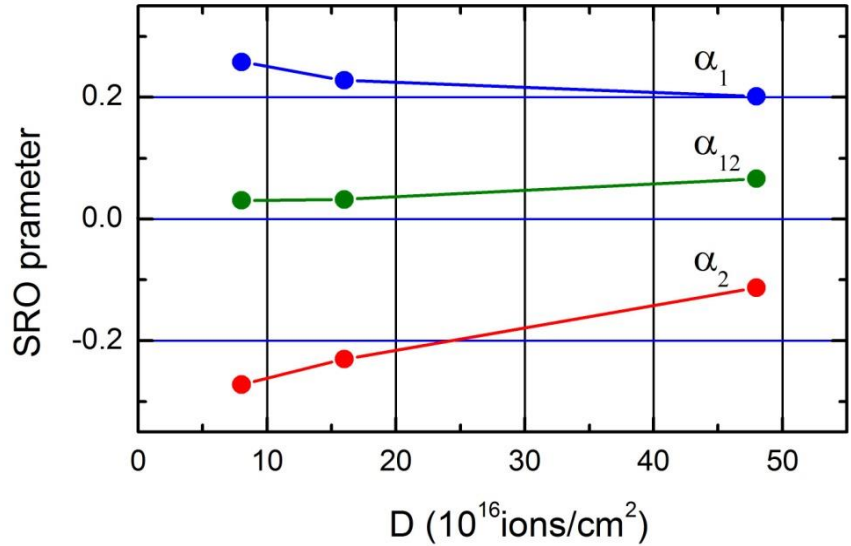

Fig. 3. SRO-parameters $\alpha_1$, $\alpha_2$ and $\alpha_{12}$ determined from the CEMS spectra recorded on the NIR-sides of the IR-samples. Typical error $\pm 0.05$. The solid lines are to guide an eye.

It follows from Fig. 3 that the SRO-parameters are characteristic of a given neighbour shell and, for a given shell, they depend on the irradiation dose, $D$. The values of $\alpha_1$ are positive what implies that the number of Cr atoms in the 1NN-shell is lower than that expected for a random distribution. This can be interpreted as evidence that the effective interaction between Fe and Cr atoms, being the nearest neighbors to each other, is repulsive. On the contrary, $\alpha_2$ has negative values thus indicating that the number of Cr atoms in the 2NN shells is higher than that expected from the binomial distribution. In other words, the effective interaction between Fe atoms and Cr atoms separated by the distance equal to the radius of the 2NN shell is attractive. Values of $\alpha_{12}$ are weakly positive what means that the number of Cr atoms within the 1NN-2NN volume around the probe Fe atoms is slightly lower than expected for the statistical distribution. The unexpected dependence of $\alpha_1$ and $\alpha_2$ SRO-parameters on the dose, D, (hence, consequently that of $\alpha_{12}$) as displayed in Fig.3 can be understood in terms of a chemical inhomogeneity of the studied samples. The inhomogeneity could be present already in the bulk alloy or it could be created by the cold rolling which was applied in order to produce thin foils used in the present study.

### 3.1.2. Irradiated (IR) side

Evaluation of the spectra recorded on the IR-sides of the samples yielded the SRO-parameters that are illustrated in Fig. 4.

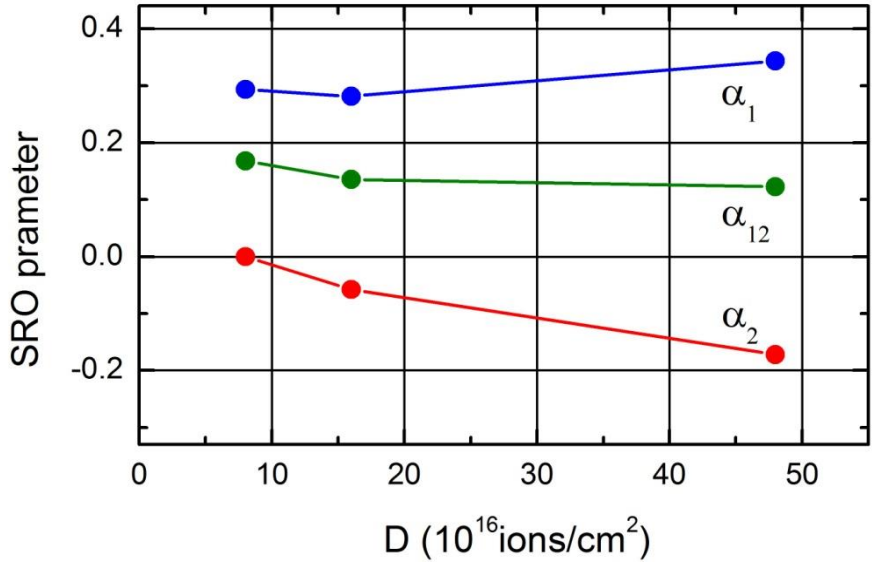

Fig. 4. SRO-parameters $\alpha_1$, $\alpha_2$ and $\alpha_{12}$ determined from the CEMS spectra recorded on the IR-sides of the samples. Typical error ±0.05. The solid lines are to guide an eye.

As in the case of the NIR-data, also in this case the SRO-parameters are characteristic of the neighbour shell. Those determined for 1NN are positive showing first a light decrease and then an increase with the dose; $\alpha_2= 0$ for D=8·10$^{16}$ions/cm$^2$ and it strongly decreases with $D$. Finally, $\alpha_{12}$ are positive for all irradiation doses but its amplitude shows a weak decrease with $D$. To reveal the real effect of the irradiation on the distribution of Cr atoms one has to take into account the initial distribution i.e. prior to the irradiation.

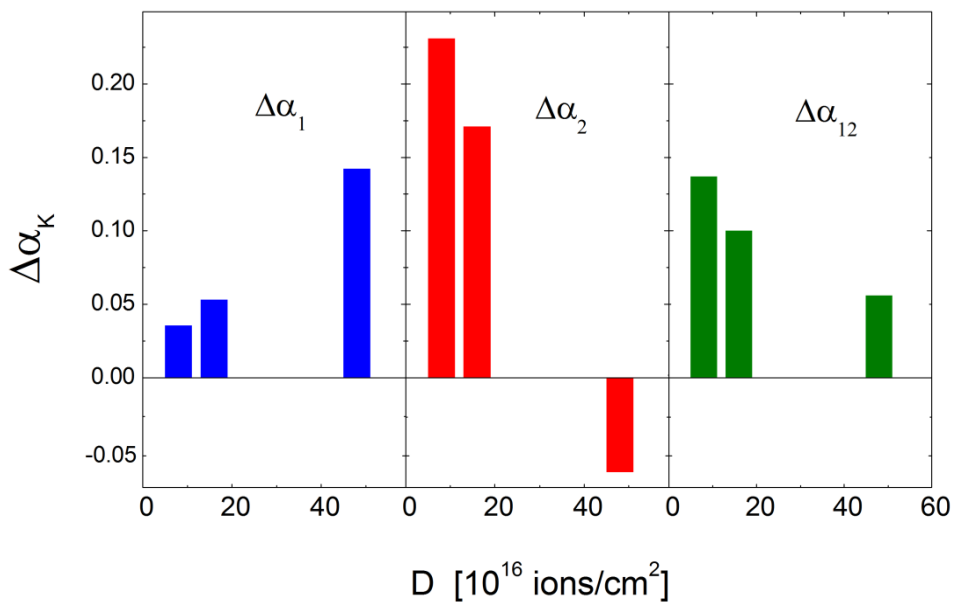

Fig. 5. Difference of the SRO-parameters $\Delta\alpha_k=\alpha_k(IR)-\alpha_k(NIR)$ (k=1,2,12) as determined from the data shown in Figs. 3 and 4.

For that purpose a difference $\Delta\alpha_k = \alpha_k(IR)-\alpha_k(NIR)$ (k=1,2,12) was calculated and presented in Fig. 5. The behavior is clearly characteristic of the neighbor shells viz. the degree of Cr atom clustering increases with the dose, D, for 1NN, while the opposite trend is seen for 2NN. The degree of the clustering averaged over the 1NN–2NN shells decreases, however, with the dose.

Precipitation of Cr-rich clusters ($\alpha'$) was revealed with ATP method in neutron-irradiated Fe-rich Fe-Cr samples [11,12]. The Cr-content in these precipitates was in the range of ~55±3 at% Cr [11] and ~85±3 at% Cr [12]. The former would be at room temperature weakly magnetic while the latter paramagnetic. If their relative contribution were ≥~1 % they would be detectable with MS. Our CEMS spectra recorded on IR-samples do not show any signs of such precipitates.

### 3.2. Local concentration change

An alternative way of describing the effect of the irradiation on the distribution of Cr atoms in the studied alloy is in terms of the underlying changes in the local Cr concentration. For that purpose we recalculate the average number of Cr atoms in 1NN shell, $<n_1>$, that in 2NN shell, $<n_2>$, as well as the one in 1NN-2NN shells, $<n_{12}>$ into the concentration, $x_k$, by using the following equation:

$$x_k(at\%) = \frac{<n_k>}{M}100 \qquad (2)$$

where $M$=8, 6, 14 for $k$=1, 2, 12, respectively.

Table 2. Concentration of chromium in the $k$-th neighbor shell, $x_k$, as calculated using eq. (2) for the non-irradiated (NIR) and irradiated (IR) sides of the samples. The dose is indicated by $D$.

| D | | NIR | | | IR | | |
|---|---|---|---|---|---|---|---|
| ions/cm² | dpa | $x_1$ | $x_2$ | $x_{12}$ | $x_1$ | $x_2$ | $x_{12}$ |
| 8·10⁶ | 5 | 11.25 | 19.3 | 14.7 | 10.7 | 15.15 | 12.6 |
| 16·10⁶ | 10 | 11.7 | 18.6 | 14.7 | 10.9 | 16.0 | 13.1 |
| 48·10⁶ | 30 | 12.1 | 16.9 | 14.15 | 9.95 | 17.75 | 13.3 |

From the data displayed in Table 2 it is evident that the concentration of Cr atoms both in NIR and in IR sides of the samples in the 1NN shell is lower than the value of 15.15 at% determined by the chemical analysis, and that in the 2NN shell it is higher. In the NIR sides of the samples the highest difference was found for the sample irradiated to the dose of $8 \cdot 10^{16}$ ions/cm$^2$ and the lowest one for the sample irradiated to the maximum dose. The concentration averaged over both neighbour shells, $x_{12}$, is, however, quite close to the value of 15.15 at%, the maximum deviation being 1 at%. This means that already in the non-irradiated alloy there was some degree of clustering and chemical inhomogeneity. In the IR sides of the samples the situation is similar as far as the relationship between $x_1$ and $x_2$ is concerned i.e. $x_1 < x_2$, yet the values of the corresponding concentrations are significantly lower for the IR sides. Consequently, the $x_{12}$-values determined for the IR sides of the samples are smaller, on average by 1.5 at%, than those found for the NIR sides. This clearly proves that the applied irradiation caused a clustering of Cr atoms.

### 3.3. Effect on the hyperfine field

Further evidence in favor of the irradiation-induced clustering of Cr atoms is shown in Fig.6. As shown the average hyperfine field, $<B> = \sum_{n_1,n_2} B(n_1,n_2) P(n_1,n_2)$, is linearly correlated with the average number of Cr atoms within the 1NN-2NN volume around the probe Fe nuclei. The <B>-values determined for the IR-samples have significantly higher values than those calculated for the NIR-samples. The observed increase of <B> by ~0.9 T is equivalent, in the light of the B(x) relationship [8], to a decrease of x by ~3(1) at%. The latter agrees reasonably with the corresponding figure deduced in paragraph 3.2.

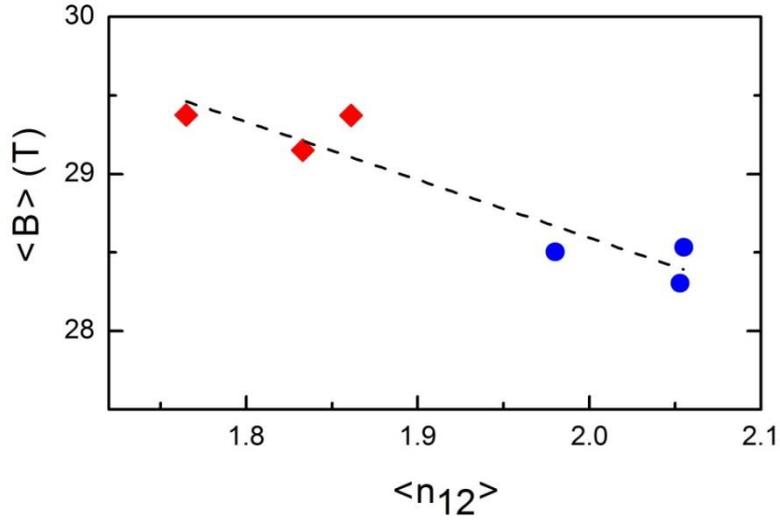

Fig. 6. Relationship between the average hyperfine field, <B>, and the average number of Cr atoms in the 1NN-2NN shells, $n_{12}$, as found for the irradiated (diamonds) and non-irradiated (circles) samples. The dashed line shows the linear correlation.

### 3.4. Effect on the magnetization vector

The knowledge of the Clebsch-Gordan coefficient $C2$ permits determination of an angle between the direction of the γ-rays (in this case perpendicular to samples' surface) and that of the magnetization vector, Θ, based on the following equation:

$$C2 = \frac{4\sin^4\Theta}{1+\cos^2\Theta} \qquad (3)$$

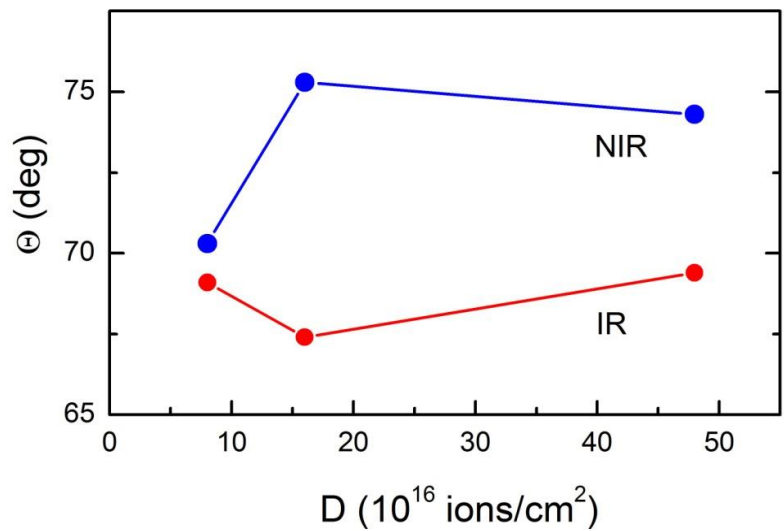

Fig. 7. Dependence of the angle Θ, on the dose of irradiation, *D*, for the non-irradiated (NIR) and irradiated (IR) sides of the investigates samples. The solid lines are a guide to an eye.

As illustrated in Fig. 7, the values of Θ in the irradiated samples are significantly smaller than those in the non-irradiated samples. It is also clear that in the NIR-samples Θ had different values reflecting thereby a lack of homogeneity as far as a cold-rolled-induced texture in the studied samples is concerned. Thus to reveal a genuine effect of the irradiation on Θ, a difference between $Θ_{IR}$ and $Θ_{NIR}$, ΔΘ, was calculated and plotted in Fig. 8.

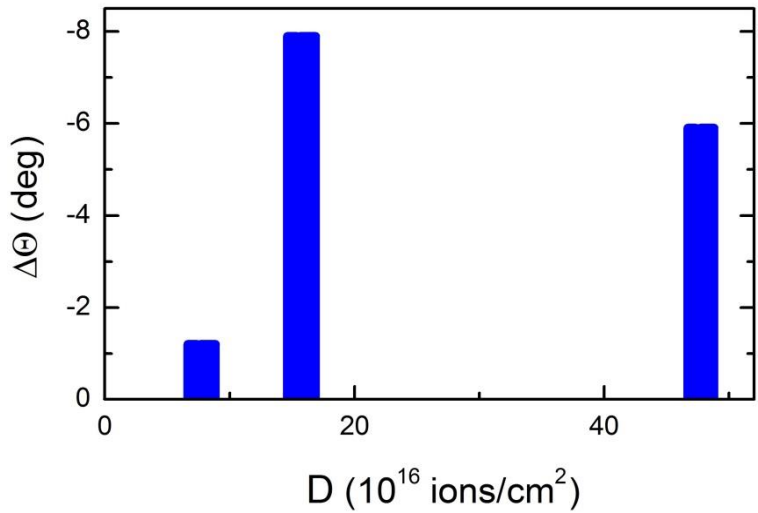

Fig.8. The difference in angle, $ΔΘ=Θ_{IR}-Θ_{NIR}$, for different doses of irradiation i.e. 8, 16, and $48·10^{16}$ ions/cm$^2$.

The data presented in Fig. 8 give evidence that Θ was different before the irradiation and it also depends on the irradiation dose. To account for the different initial values of Θ, a difference $ΔΘ=Θ_{IR}-Θ_{NIR}$ was calculated. The maximum decrease of Θ occurred for the sample irradiated to $D=16·10^{16}$ions/cm$^2$ and the minimal one for the one irradiated to $8·10^{16}$ions/cm$^2$. The decrease of Θ upon irradiation of the $Fe_{85}Cr_{15}$ alloy with 25 keV He ions was already observed [10]. However, in that case the maximum change of Θ was ~13°. The significantly higher rotation in the latter could have its origin in a much lower implantation range viz. 0.2 μm for 25 keV ions than that of 0.6 μm for 250 keV ions. Consequently, the rotation of the magnetization vector in the samples irradiated with 25 keV ions was due not only to a ballistic effect, as in the case of 250 keV ions, but also to a presence of He atoms in the pre surface layer accessible to CEMS measurements.

## 4. Conclusions

The results obtained in this study permit the following conclusions to be drawn:

1. The distribution of Cr atoms in a heavily deformed $Fe_{85}Cr_{15}$ alloy is not random: the nearest-neighbor shell is underpopulated ($\alpha_1<0$) while the second nearest-neighbor shell is overpopulated ($\alpha_2>0$) with Cr atoms.

2. The irradiation with 250 keV $^4He^+$ ions caused redistribution of Cr atoms: their number decreased in 1NN, and increased in 2NN with the dose, D. The degree of the clustering determined within the 1NN-2NN volume was revealed to decrease with D.

3. The irradiation induced redistribution resulted in: (a) clustering of Cr atoms reducing thereby the local concentration of Cr atoms within the 1NN-2NN vicinity of Fe atoms by 0.85 - 2.1 at%, (b) increase of the average hyperfine field by ∼0.9 T, and (c) rotation of the magnetization vector towards the normal to the samples' surface by maximum of ∼8° for the sample irradiated to the dose of $16·10^{16}$ ions/cm.

**Acknowledgement**

This work has been carried out within the framework of the EUROfusion Consortium and has received funding from the European Union's Horizon 2020 research and innovation program under grant agreement number 633053. The views and opinions expressed herein do not necessarily reflect those of the European Commission. It was also supported by The Ministry of Science and Higher Education, Warszawa, Poland. Mr. Yves Surreys is thanked for his expert assistance in the irradiation.